\documentclass[prd,
eqsecnum,floatfix,letterpaper,showpacs,superscriptaddress,groupedaddress,nofootinbib]{revtex4}

\usepackage{amsmath}
\usepackage{latexsym}
\usepackage{amssymb}
\usepackage{amsfonts}
\usepackage{mathtools}
\usepackage{bm}
\usepackage{amsthm}
\usepackage{graphicx}
\usepackage{color}
\usepackage{changes}
\usepackage[normalem]{ulem}
\usepackage{natbib}
\usepackage{appendix}
\def \D{{\mathcal D}}
\def \G{{\cal G}}
\def \a {\alpha}
\def \b {\beta}
\def \g {\gamma}

\def \V {V^{\rm trans}}
\def \B {B^{\rm trans}}
\def \P {P_{\rm interleave}}
\def \tpsi {\tilde{\psi}}
\def \tpsi {\tilde{\psi}}
\def \tP {\tilde{P}}
\def \tB {\tilde{B}}
\def \tS {\tilde{S}}
\def \tT {\tilde{T}}

\def \be{\begin{equation}}
\def \bea{\begin{eqnarray}}
\def \eea{\end{eqnarray}}
\def \ee{\end{equation}}

\begin{document}
\title{The Varied Avatars of Time-delay Interferometry }
\author{Sanjeev Dhurandhar}
\email{sanjeev@iucaa.in}
\affiliation{Inter University Centre for Astronomy and Astrophysics,
  Ganeshkhind, Pune, 411 007, India}
\author{Prasanna Joshi}
\email{prasanna.mohan.joshi@aei.mpg.de}
\affiliation{Max Planck Institute for Gravitational Physics (Albert-Einstein-Institute), D-30167 Hannover, Germany}
\author{Massimo Tinto}
\email{mtinto@ucsd.edu}
\affiliation{University of California San Diego,
  Center for Astrophysics and Space Sciences,
  9500 Gilman Dr, La Jolla, CA 92093,
  U.S.A.}
\affiliation{Divis\~{a}o de Astrof\'{i}sica, Instituto
  Nacional de Pesquisas Espaciais, S. J. Campos, SP 12227-010, Brazil}
\date{\today}

\begin{abstract}
  Time-Delay Interferometry (TDI) is the data processing technique
  that cancels the large laser phase fluctuations affecting the
  one-way Doppler measurements made by unequal-arm space-based
  gravitational wave interferometers. By taking finite linear
  combinations of properly time-shifted Doppler measurements, laser
  phase fluctuations are removed at any time $t$ and gravitational
  wave signals can be studied at a requisite level of sensitivity.
 
  In the past, other approaches to this problem have been
  proposed. Recently, matrix based approaches have been put forward;
  two such approaches are by Vallisneri {\it et al.} and Tinto,
  Dhurandhar and Joshi. In this paper we establish a close
  relationship between these approaches. In fact we show that the
  matrices involved in defining the operators in the two approaches
  exhibit an isomorphism and therefore in both approaches one is dealing with matrix representations of the time-delay operators. 
\end{abstract}

\maketitle

\section{Introduction}
\label{SecI}

In ground based detectors of gravitational waves (GWs) the arms are
chosen to be of equal length. This is because
the laser phase fluctuations experience identical delays in
the arm of the interferometer and cancel at
  the photo-detector where the two returning beams are made to
  interfere. In space-based detectors, on the other
hand, the arm-lengths are unequal and time-dependent as each
  spacecraft follows a trajectory determined by celestial mechanics.
As a result it becomes impossible to maintain the distances
  between spacecraft equal and constant. Time-Delay Interferometry (TDI) is required to cancel
the laser phase noise, which is many orders of magnitude
above the other residual noise sources (such as shot
noise, test mass acceleration noise, etc.) affecting the
  heterodyne one-way measurements. TDI entails properly
delaying and linearly combining the
different data streams so that the laser phase fluctuations are suppressed below the residual noises and GW signals may be observed. 

  In the past, other approaches have been proposed to compensate for
  the inequality of the arms and achieve suppression of the laser
  noise below the residual noise levels. The first, which was formulated
  in the Fourier domain \cite{GHTF96}, represented the delayed one-way
  measurements in terms of their Fourier transforms multiplied by
  corresponding phasors. This approach was incorrect for two
  fundamental reasons. Firstly, it neglected the time-evolution of the
  delays, which we now know needs to be accounted for to sufficiently
  suppress the laser noise. Secondly (and more importantly), it made
  the erroneous assumption of taking infinitely long Fourier
  transforms of the delayed one-way measurements. A finite-time
  Fourier transform of a delayed measurement is not equal to the
  product of its Fourier transform with the delay phasor. Rather it is
  equal to the Fourier domain convolution of the Fourier transform of
  the data with the Fourier transform of the window of
  integration. This implied the existence of residual laser noise
  terms in the Fourier-domain laser-noise canceling algorithm that
  could be neglected only by taking 6 months or longer Fourier
  transforms of the measurements \cite{TA99}.

A much neater approach by Tinto, Estabrook and Armstrong, where the
delays were represented by derivative-like symbols, namely commas
\cite{TA99,AET99,TEA02,TEA04}, was applied in the context of the 
LISA mission \cite{LISA98,LISA2017}. This notation (and its
understanding) facilitated the algebra of the TDI observables and led
in principle, to a plethora of such observables that could be obtained
conveniently by just linearly combining the four Sagnac observables 
$\a, \b, \g, \zeta$. This work showed that the space of TDI was a
linear object and its elements could be obtained by linearly combining four simple basic TDI observables. This is in fact a
most important result of this approach.

An exceptionally deep insight into time-delay interferometry  was obtained, when Dhurandhar, Nayak
and Vinet \cite{DNV02} found the exact underlying mathematical
structure of the TDI space. In this approach the time-delay operation
was promoted to operators acting on data streams or operators
acting on functions of time. The operators played the role of
indeterminates in a polynomial ring and the TDI space was none other
than the first module of syzygies \cite{becker}. It was shown therein that the TDI
space is a module over the polynomial ring of time-delay operators and
hence pinned down the linear structure. This work laid emphasis on the
operators rather than on the functions (data streams containing laser
noise) since these were subsidiary - the function space is the carrier
space. This is a similar situation as one has in matrix
representations of groups; the matrices, which are linear maps on
the carrier vector space, represent the group elements. This is
interesting from the historical point of view because the first module
of syzygies was defined by Hilbert in 1890 in a different context. It
is in fact a kernel of a homomorphism. This is exactly what one
desires - its physical significance here is that elements in a kernel
map to zero. In the current context this is the zero of
the laser noise:  we are looking for those data
combinations that map the laser phase noise to
zero. This approach rigorously proved that all TDI observables can be
obtained as a linear combination of the four generators
$\a, \b, \g, \zeta$ \footnote{A module in general does not have a
  basis, but has generators which span the module though they may not
  be linearly independent - one may not be able to reduce the number
  of generators in general, because multiplicative inverses need not
  exist in a ring.}.  The first work \cite{DNV02} considered constant
arm-lengths. Later more general and realistic models of LISA were
considered \cite{NV04,DNV08,DNV10,STEA03,TEA04} which increased the
complexity. These approaches with generalizations have been reviewed
comprehensively in \cite{TD2020}.

In fairly recent years a novel approach was adopted by Vallisneri {\it
  et al.}  \cite{Vallisneri2020} in which the problem has been formulated
in terms of matrices (see also \cite{PCA2006} where principal
component analysis has been employed). In this approach, the data are
discretized and a design matrix representing the delays is
defined. But since data points may be required in between the sample
points for TDI to be effective, an interpolating scheme must be
employed for fractional delays. Here also a null space is sought whose
elements are then the TDI observables. Another matrix approach was put
forward by Tinto, Dhurandhar and Joshi\cite{TDJ21}. In this work, it
was shown that the matrix approach is a ring representation
\cite{Burrow} of the operator approach - there is a homomorphism
between the ring of operators into the ring of matrices. However, the
matrix approach seems to have an advantage because matrices are easy
to manipulate (although this has not been conclusively
established). This is in the same spirit, as one uses group
representations rather than abstract group elements to perform
calculations. In this paper we show that there is in fact an
isomorphism (which is more than homomorphism - the map is also one to
one and onto) between the design matrices defined in
\cite{Vallisneri2020} and the matrix operators defined in our approach
\cite{TDJ21}. 
\par

This paper is organized as follows. In Sec. \ref{SecII} we describe the
design matrices defined in \cite{Vallisneri2020} but extract half the
rows corresponding to one of the arms because they have the basic
structure we want to investigate.  In Sec. \ref{isomorphism} we then
prove the isomorphism between the matrices defined in \cite{TDJ21} and
\cite{Vallisneri2020}, while in Sec. \ref{twoarms} we show how to
generalize the one-arm results to the two-arms configuration. In
Sec. \ref{conclude} we finally present our concluding remarks, and
emphasize that the isomorphism existing between our matrix
representation of the TDI delay operators and the matrices introduced
in \cite{Vallisneri2020} should help us in relating the laser
noise-free combinations identified by the two methods.

\section{Algebra of design matrices: the case of the single arm} 
\label{SecII}

 In this section, we will consider just one of the arms, say arm 1, that is, we do not consider the interleaving of $y_1$ and $y_2$ discussed in \cite{Vallisneri2020}. In section \ref{twoarms} we will indicate how to generalise the analysis to the case for two arms. Also we will only consider that part of the matrix which describes the delay and therefore disregard the subtraction of the unit matrix; that is, if $M$ is the design matrix for one arm, then we consider the matrix $V = M + I$, where $I$ is the identity matrix. We denote the matrix by $V$  since the design matrix has been introduced in the paper by Vallisneri et al \cite{Vallisneri2020}. Also for simplicity (as mentioned in the same paper) we will take the sampling interval to be unity. The delayed data are given by $\D_1 y(t) = y (t - l_1)$.  It is first of all easily verified that for integer delays the product of two operators say $\D_1$ and $\D_2$ corresponds to the product of the matrices $D_1$ and $D_2$. These delay matrices for delays $\Delta t = 1, 2, 3$ have been explicitly displayed in \cite{TDJ21}. The $V$ matrices for integer delays can be taken to be  identical to the $D$ matrices and so the homomorphism for the $V$ matrices follows as shown in \cite{TDJ21}. This disposes of the integer delays.
\par

For fractional delays, we only need to exhibit a bijective map between the matrices $D$ defined in \cite{TDJ21} and the matrices $V$. We do so below. For this purpose, we will first need to describe the structure of the matrix $V$ for fractional delays. The Lagrange polynomial interpolation scheme is chosen as in \cite{Vallisneri2020} with the degree $m$ of the polynomials set equal to $6$. The data $y$ are labelled at integer nodes at $t_k = k,~ k = 0, 1, 2, ...$ and are denoted accordingly by $y_k = y (t_k)$. Now consider a fractional delay $\a$. The non-integer delay $\a$ is broken up into its integer part $[\a]$ and the residual part $\delta \a$ accordingly as $\a = [\a] + \delta \a$. Now the interval containing $m$ nodes has to be chosen so that it covers the delayed time instant and such that it lies somewhere near the centre of the interval. This depends at what time instant we are evaluating the delayed data. If the time instant is $k$, then we go back $[\a] + m/2 = [\a] + 3$ nodes, and it is at this node the filter mask starts. For example, if $\a = 2.2$, then $[\a] = 2$ and one must start the interpolating interval from $k - 5$. If the data are measured from $t = 0$, then a full mask is possible only when $k \geq 5$. The first such instant occurs at $k = 5$ and the interval is $\{0, 1, 2, 3, 4, 5 \}$ with node values $\{y_0, y_1, ..., y_5 \}$. The interpolated data is evaluated at $t = 5 - \a = 2.8 \equiv \a'$ say, or $y(2.8)$. For the sake of completeness we give below the general expressions for the Lagrange polynomials on $m$ nodes at $t_k,~ k = 0, 1, ...,m - 1$:
\be
l_k (t) = \frac{(t - t_0)(t - t_1) ... (t - t_{k - 1})(t - t_{k + 1}) ...(t - t_{m - 1})}{(t_k - t_0) ...(t_k - t_{k-1})(t_k - t_{k+1})...(t_k - t_{m - 1})} \,, ~~~~k = 0, 1, ...,m - 1.
\ee
Note that $l_k (t)$ are polynomials of degree $m - 1$. Here we have $m = 6$ and so the polynomials are of degree 5. Then the interpolated value of $y$ at $t = \a'$ is given by,
\be
y (\a') = \sum_{k = 0}^5 l_k (\a') y_k \,.
\ee
In the $V$ matrix, the first row with 6 non-zero entries occurs first at the 5$^{\rm th}$ row. Then we have $V_{5k} = l_k (\a')$ for $0 \leq k \leq 5$ and $V_{5k} = 0$ for $k > 5$. In the next row $k = 6$, the filter mask covers the interval $\{1, 2, ..., 6 \}$ and the corresponding data points are $\{y_1, y_2, ..., y_6 \}$. In the matrix $V$, the Lagrange polynomials are shifted by one column to the right with $V_{60} = 0$ and $V_{6k} = 0$ for $k = 1, 2, ..., 6$. Here for simplicity we have chosen the time-delay to be constant (the time dependent case does not make much difference to the homomorphism  argument).  As one proceeds down the rows, the Lagrange polynomials get shifted to the right and so diagonally downwards. The matrix $V$ looks as follows:
\begin{equation}
V (\a') = \left(
    \begin{array}{cccccccccc}
         \vdots & \vdots & \vdots & \vdots & \vdots & \vdots & \vdots & \vdots & \vdots & \cdots \\
         l_0 (\a') &  l_1 (\a') &  l_2 (\a') &  l_3 (\a') & l_4 (\a')  & l_5 (\a') &  0 & 0 & 0 & \cdots \\
         0 & l_0 (\a') &  l_1 (\a') &  l_2 (\a') &  l_3 (\a') & l_4 (\a')  & l_5 (\a') &  0 & 0 & \cdots   \\
         0 & 0 & l_0 (\a') &  l_1 (\a') &  l_2 (\a') &  l_3 (\a') & l_4 (\a')  & l_5 (\a') &  0 & \cdots  \\
\vdots & \vdots & \vdots & \vdots & \vdots & \vdots  & \vdots & \vdots & \vdots & \ddots
    \end{array}
  \right) 
  \end{equation}
  The first non-trivial row shown is for $k = 5$. For $\a' = 2.8$, the $l_k$ take the numerical values $0.006336, -0.04928, 0.22176, \break 0.88704, -0.07392, 0.008064$ as $k$ ranges from $0$ to $5$ in steps of unity. 
  
  \section{The isomorphism between $V$ and $D$ matrices}
  \label{isomorphism}
  
  For establishing the homomorphism this arrangement presents difficulties because the target subspace, namely, the carrier space, changes (advances) with each successive row. For $k = 5$ the target subspace is the interval $W_0 = \{0, 1, 2, 3, 4, 5 \}$ while for $k = 6$, the target subspace is $W_1 = \{1, 2, 3, 4, 5, 6 \}$ and so on. In order to establish homomorphism one requires a fixed target subspace. We therefore fix a target subspace. There are several choices for this; we make the following one. We fix the target subspace to be $W_0$ and so refer all the Lagrange polynomials to $W_0$; in effect we translate the Lagrange polynomials to $W_0$. This means, for example, the entries in the 6$^{\rm th}$ row ($k = 6$), need to be shifted by one column to the left. This is achieved by translating the Lagrange polynomials - that is by adding $1$ to the argument $\a'$. This makes sense because for the value of $\a' = 2.8$, we evaluate the Lagrange polynomial at $\a' + 1 = 3.8 = 6 - 2.2 = 6 - \a$. Thus the translated Lagrange polynomials are $l_k (\a' + 1)$.  Similarly, for the next row $k = 7$, one needs to shift the Lagrange polynomials by two columns to the left and therefore we must add 2 to the argument $\a'$ resulting in $l_k (\a' + 2)$. Thus the entries in rows $k = 5, 6, 7, 8, 9, 10$ are shifted to the left by the  appropriate number of columns with the arguments of the Lagrange polynomials increased by the number equal to the number of shifted columns. By carrying out this procedure we obtain the translated matrix $\V$ given below:
  
\begin{equation}
\V (\a') = \left(
    \begin{array}{cccccccccc}
         \vdots & \vdots & \vdots & \vdots & \vdots & \vdots & \vdots & \vdots & \vdots & \cdots \\
         l_0 (\a') &  l_1 (\a') &  l_2 (\a') &  l_3 (\a') & l_4 (\a')  & l_5 (\a') &  0 & 0 & 0 & \cdots \\
         l_0 (\a' + 1) &  l_1 (\a' + 1) &  l_2 (\a' + 1) &  l_3 (\a' + 1) & l_4 (\a' + 1)  & l_5 (\a' + 1) &  0 & 0 & 0 & \cdots   \\
         l_0 (\a' + 2) &  l_1 (\a' + 2) &  l_2 (\a' + 2) &  l_3 (\a' + 2) & l_4 (\a' + 2)  & l_5 (\a' + 2) &  0 & 0 & 0 & \cdots  \\
\vdots & \vdots & \vdots & \vdots & \vdots & \vdots  & 0 & 0 & 0 & \cdots   \\    
l_0 (\a' + 5) &  l_1 (\a' + 5) &  l_2 (\a' + 5) &  l_3 (\a' + 5) & l_4 (\a' + 5)  & l_5 (\a' + 5) &  0 & 0 & 0 & \cdots  \\
\vdots & \vdots & \vdots & \vdots & \vdots & \vdots  & \vdots & \vdots & \vdots & \ddots 
    \end{array}
  \right) 
\end{equation}
We immediately recognise that the above block matrix of the Lagrange polynomials is identical with the $D$ matrix, namely,  Eq. (4.13), of reference \cite{TDJ21}. Since as shown in there, the $D$ matrices form a representation of the fractional delay operators, it follows that the block matrices in $\V$ also constitute a representation of the same operators. 
\par

We can also perform the operation of shifting the rows. This is just translating the polynomials by the required time stamps. If we shift the rows 'upwards' by $r$ time samples, then we must subtract $r$ from the argument of the Lagrange polynomials. For example, in the above example, if we shift by 2 time samples upwards, the arguments in any column of the $6 \times 6$ block will range from $\a' - 2$ to $\a' + 3$.  Thus in our case of $\a' = 2.8$, the arguments will range from $0.8$ to $5.8$ close to the interpolation nodes of $W_0$. This is relevant when numerical accuracy is a consideration. The opposite happens if we shift down and to the right - shifting diagonally downwards and to the right (or upwards and to the left) keeps the entries in each row the same - the arguments of the Lagrange polynomials do not change. We have assumed here constant armlengths for simplicity. We will remark later on time-dependent armlengths; they do not cause any difficulty in principle to the homomorphism argument.   
\par

The above discussion was relevant to the first $6 \times 6$ block. The next $6 \times 6$ block begins at the 11$^{\rm th}$ row, that is, $j = 11$ to $j = 16$ and the columns are from $k = 6$ to $k = 11$. Thus the target subspaces start from $W_6$ onwards. We may now fix the target subspace to be $W_6$ and repeat the steps above. This will lead to an identical block matrix as the first one. Thus the matrix represention of $\V$  is block diagonal, each block a $6 \times 6$ matrix.  
\par

We can describe this representation directly in general for a filter mask on $m$ nodes. Let $\a$ and $\b$ be two time delays. Define $r = [\a] + m/2$ and $s = [\b] + m/2$, then $\V_{r + j, k} (\a') = l_j (\a' + k)$ and $\V_{s + k, n} (\b') = l_k (\b' + n)$ then:
\be
\sum_{k = 0}^{m - 1} \V_{r + j, k} (\a') \V_{s + k, n} (\b') = \V_{r + s + j, n} (\a' + \b') \,.
\label{hom}
\ee
The above equation follows from the addition theorem for Lagrange polynomials. Eq. (\ref{hom}) directly exhibits the homomorphism. This is in fact the same as equation Eq. (4.15) given in reference \cite{TDJ21}. Since $m/2$ is added both to $r$ and $s$, we may shift the rows by $m/2$ upwards by writing the RHS of Eq. (\ref{hom}) as $\V_{r + s + j - m/2, n} (\a' + \b' - m/2)$ to get it in the required form as given in \cite{Vallisneri2020}. This rule obtains if $\delta \a + \delta \b < 1$. If this is not the case, that is, if $\delta \a + \delta \b \geq 1$ then the required upward shift is $m/2 - 1$.
\par

The time dependent case follows exactly the discussion in \cite{TDJ21}. If the delay $\b$ is applied after $\a$, then $\b$ becomes a function of $\a$ and  the composite delay is given by $\a + \b (\a)$. If the order of the delays is reversed, then the composite delay is $\b + \a (\b) \neq \a + \b (\a)$. Thus the delay operators and their representative matrices do not commute in general. This is the basic difference between the time dependent and time independent cases. Rest of the discussion parallels the discussion for the time independent case.
\par

The only point remaining is to formally establish the correspondence between $V$ and $\V$. Let this correspondence be denoted by the mapping $\psi$. We show below that $\psi$ is in fact an isomorphism (not merely a homomorphism). Thus we have,
\be
\psi [V (\a')] = \V (\a') \,.
\ee
The above discussion shows that:
\be
\psi [V (\a' + \b')] = \V (\a' + \b') = \V (\a') \star \V (\b') = \psi [V (\a')] \star' \psi [V (\b')] \,,
\label{intertwine}
\ee
where the $\star$ and $\star'$ operations are defined as  above.
\par

Here some remarks are in order. It is first of all clear that $\psi$ is linear and hence a homomorphism. Eq. (\ref{intertwine}) in fact defines the operation $\star'$. We will establish below that $\psi$ is bijective and hence an 
\emph{ isomorphism}. We will explicitly establish this fact by exhibiting formulae. In the literature \cite{Burrow}, the map $\psi$ or more appropriately its extension $\tpsi$ is called an \emph{intertwinor} and the representation an intertwining representation. Thus the matrices $V$ form an intertwining representation. In the appendix \ref{app:intertwine} we  indicate how the map $\psi$ is extended to the intertwining map $\tpsi$.
\par

Let us consider first, the first $6 \times 6$ block of $V$. It is easily shown that the $6$ row vectors of the block matrix, namely, $R_k, ~k = 0, 1, ...,5$ are linearly independent and therefore span a $6$ dimensional subspace $B_1 = R_0 \oplus R_1 \oplus R_2 \oplus ...\oplus  R_5$, where the symbol $\oplus$ denotes direct sum. $B_1$ is the domain of the map $\psi$. Now we come to the range of $\psi$ which is the first block $\B_1$ of $\V$. $\B_1$ also consists of 6 linearly independent row vectors, because the inverse $[\V (\a)]^{-1} = \V (-\a)$ in principle always exists - we can always undo the delay by reversing the situation. The existence of the inverse implies that the row space of $\B_1$ must be $6$ dimensional. This proves that the map $\psi$ is bijective where $\psi (B_1) = \B_1$. This map can be extended in an obvious way to the rest of the blocks of $V$ and $\V$, hence establishing formally the correspondence as desired. 
\par

The isomorphism can also be proved explicitly by computing the translation matrices. We show this only for the block $B_1$. The argument extends in an obvious way to $V$. But before we apply the translation matrices, we need to shift the rows of $B_1$ to the left by the appropriate number of columns. This is done by projecting out each row at a time by applying the projection operators $P_k, ~k = 0, 1, ..., 5$ and then shifting to the left by the required number of columns. The matrices $P_k$ are $6 \times 6$ and have all entries $0$, except for $1$ on the $k^\text{th}$ row and column. Applying $P_k$ on the left of $B_1$ picks out the $k^\text{th}$ row, zeroing out other rows. Then we need to shift the rows by $k$ columns to the left. This is achieved by applying shift matrices $S_k$ which are $11 \times 6$. These are nothing but essentially the delay operators $D_k$ for integer valued delays. Finally, we apply the translation matrices $T_k$ which map a row vector $L_0 \longrightarrow L_k$, where we define the row vector $L_k = [l_0(\a + k, l_1 (\a + k), ..., l_5 (\a +k)]$. Thus, we write $L_k = L_0 T_k$. The $T_k$ are $6 \times 6$ matrices. We therefore obtain:
\be
\B_1 = \sum_{k = 0}^5 P_k B_1 S_k T_k \,.
\label{psi}
\ee
This is the map $\psi$. 
\par

We can also invert the above relation. We only need to multiply the above sum by the projection operator $P_j$ from the left; because $P_j P_k = P_j \delta_{jk}$. This picks out the $j^\text{th}$ term zeroing out the other terms. Thus we obtain:
\be
B_1 = \sum_{k = 0}^5 P_k \B_1 T_k^{-1} S_{-k} \,.
\label{psiinv}
\ee
This is the inverse map $\psi^{-1}$. Here $S_{-k}$ is the operator which shifts the elements to the right by $k$ columns and undoes the effect $S_k$. Further, the $T_k$ matrices are invertible - in fact $\det{T_k} = 1$. We can compute them explicitly. We can write any $l_n (\a + k)$ as linear combinations of $l_j (\a)$. For example, $l_0 (\a + 1) = - l_5 (\a), l_1 (\a + 1) = l_0(\a) + 6 l_5 (\a), ...$.  This equation can be written in matrix form: $L_1 = L_0 T_1$, where,
\be
T_1 = \left[
    \begin{array}{ccccccc}
         0 &  1 &  0 &  0 & 0  & 0  \\
         0 &  0 &  1 &  0 & 0  & 0  \\
         0 &  0 &  0 &  1 & 0  & 0 \\
         0 &  0 &  0 &  0 & 1 & 0   \\
         0 &  0 &  0 &  0 & 0 & 1 \\
         -1 &  6 &  -15 &  20 & -15 & 6 
    \end{array}
  \right] \,.
\ee
The other translation matrices $T_k$  can be obtained easily from the addition theorem for Lagrange polynomials.  All the translation matrices $T_k$ are non-singular and in fact have determinant $1$. This is because the vectors are rigidly translated, keeping the volume of the parallelepiped defined by those vectors invariant. 
\par

Thus the isomorphism can be explicitly established directly. 

\section{The generalisation to two arms}
\label{twoarms}

In this section we indicate how to generalise to the case of two unequal arms. Here we have in general two different delays $\a_1$ and $\a_2$ corresponding to the arms $1$ and $2$ respectively. The mathematical structure is that of the product ring. We will do this for a group $\G$. The product is denoted by  $\G \times \G$. It is defined as follows.
\par

Let $\G$ be a group and let $g_1, g_2 \in \G$ then the element of $\G \times \G$ is the ordered pair $(g_1, g_2)$ or we write $(g_1, g_2) \in \G \times \G$. The composition law in $\G \times \G$ is defined in the following way: Let $(g_1, g_2)$ and $(h_1, h_2)$ belong to $\G \times \G$, then $(g_1, g_2) \cdot (h_1, h_2) = (g_1 h_1, g_2 h_2)$. Clearly the product so defined is in $\G \times \G$. It is easily shown that under this composition law $\G \times \G$ is a group. 
\par

The next point to consider is a matrix representation of $\G$ which associates a matrix $T_g$ with each element $g \in \G$. The matrices $T_g$ are actually linear maps from a vector space $T_g: W \longrightarrow W$. A representation is a homomorphism $\phi$ which takes $g \in \G$ to $T_g$ or $\phi (g) = T_g$, such that for all $g, h \in \G$, $\phi (g h) = T_g T_h$ and $\phi (e) = I$ or the identity of the group $e$ is mapped to the identity matrix. 
\par

From the above considerations, we may easily define a representation of $\G \times \G$. Consider a finite dimensional representation, that is, $\text{dim} (W) = n$. Then $T_g$ is a $n \times n$ matrix. Now consider an element $(g, h) \in \G \times \G$, then we have the corresponding $n \times n$ matrices $T_g$ and $T_h$. We define the product representation $\phi \otimes \phi$ by the block diagonal $2n \times 2n$ matrix:
\be
\phi \otimes \phi [(g, h)] =  \left[
    \begin{array}{cc}
         T_g & 0  \\
         0  & T_h
    \end{array}
  \right] \,.
  \label{prod}
\ee
It is easy to show that $\phi \otimes \phi$ constitutes a representation of $\G \times \G$.  It is important to note that the two block matrices are essentially independent of each other. 
\par

For the two arm case there  are in general two independent time-delays, say $\a_1$ and $\a_2$ (these are the $g$ and $h$ in the above discussion). Under the representation homomorphism they map to $6 \times 6$ matrices $\B_1 (\a_1)$ and $\B_1 (\a_2)$ respectively or if one considers the entire matrices $\V (\a_1)$ and $\V (\a_2)$. Under the isomorphism $\psi$, $\V$ matrices map to the $V$ matrices. These can be arranged as $n \times n$ block diagonal matrices as in Eq. (\ref{prod}) to obtain the product representation  matrix which is $2n \times 2n$. Then by the following matrix transformation below, this block diagonal matrix is converted into a $2n \times n$ matrix:  
\be 
\left[ \begin{array}{cc}  V (\a_1) & 0 \\  0 & V (\a_2)  \end{array} \right] 
\left[\begin{array}{c} I_n \\ I_n \end{array} \right ] 
= \left[ \begin{array}{c} V (\a_1) \\ V (a_2) \end{array} \right]  \, ,
\label{col_shift}
\ee
where $I_n$ is the $n \times n$ unit matrix. This is not an isomorphism. Finally, the design matrix $M$ (except for the subtraction of the identity matrix) is obtained by a permutation of the rows of the matrix on the RHS of Eq. (\ref{col_shift}). The permutation operation can also be represented by a matrix $\P$ consisting of $0$ s and $1$ s. The matrix $\P$ is just $\{\delta_{2m - 1, m}, \delta_{2m, n + m} \},~~~m = 1, 2, ..., n$. It is non-singular and hence also an isomorphism. Thus we have an intimate relation between the design matrices $M$ of \cite{Vallisneri2020} and the matrices $\V$ which form a representation of the delay operators.  

\section{Concluding Remarks}
\label{conclude}

 Future space-based gravitational wave interferometers will rely on the
use of TDI to achieve their baseline sensitivities. The matrix
representations of the TDI delay operators discussed in this article
should simplify and make the implementation of TDI more efficient and
consequently the analysis of gravitational wave signals that we are searching for. In the process we have made a detailed analysis of the design matrices. In this article we have shown that the matrix representation of the delay
operators derived in \cite{TDJ21} is isomorphic (i.e. one-to-one and
onto) to the matrices introduced in \cite{Vallisneri2020} to cancel
the laser noise. The isomorphism we have just established should help
us in identifying a systematic way for relating the laser noise-free
combinations identified by TDI to those obtained by the method
proposed in \cite{Vallisneri2020}. This will be the subject of a
forthcoming investigation.
\section*{Acknowledgments}

The authors would like to thank Hemant Bhate for useful conversations on group representation theory. 
M.T. thanks the Center for Astrophysics and Space Sciences (CASS) at
the University of California San Diego (UCSD, U.S.A.) and the National
Institute for Space Research (INPE, Brazil) for their kind hospitality
while this work was done. S.V.D. acknowledges the support of the
Senior Scientist Platinum Jubilee Fellowship from NASI, India.

\appendix
\section{The interwining map $\tpsi$}
\label{app:intertwine}

Although the map $\psi$ exhibits closure property and associativity, it does not map the identity to identity because $B_1$ is not a square matrix. This is easily remedied by suitably augmenting the matrices by identity and  zero matrices. The map $\psi$ is extended to $\tpsi$ as follows. In Eqs. (\ref{psi}) and $(\ref{psiinv})$ we make all matrices $11 \times 11$. In the projection matrix $P_k$ we let $k = 0, 1, .., 10$ and call it $\tP_k$. We also define $\tB_1$, $\tS_k$, $\tT_k$ by appropriately adjoining $5 \times 5$ identity matrices and zero matrices as follows:
\be
\tB_1^{\rm trans} =  \left[
    \begin{array}{cc}
         \B_1 & 0  \\
         0  & I_5
    \end{array}
  \right] \,, ~~~
\tB_1 =  \left[
    \begin{array}{ccc}
          & B_1 &  \\
         0  && I_5
    \end{array}
  \right] \,, ~~~
  \tS_k =  \left[
    \begin{array}{cc}
          & 0  \\
          S_k & \\
          &  I_5
    \end{array}
  \right] \,, ~~~
 \tT_k =  \left[
    \begin{array}{cc}
         T_k & 0  \\
         0  & I_5
    \end{array}
  \right] \,,
\ee
where $I_5$ is a $5 \times 5$ identity matrix and the block matrices $0$ have appropriate dimensions to make all the matrices $11 \times 11$.  Then the corresponding quantities in Eq. (\ref{psi}) and $(\ref{psiinv})$ can be replaced by the quantities with tildes, and the sum over $k$ over 6 terms is replaced by a sum over 11 terms, $k = 0, 1, 2, ..., 10$. It is now easy to check that $\tpsi$ not only satisfies the properties of $\psi$ but also maps identity to identity, that is $\tpsi (I_{11}) = I_{11}$, where $I_{11}$, the $11 \times 11$ identity matrix.  

\bibliographystyle{apsrev}
\bibliography{refs}
\end{document}